\documentclass[12pt]{article}
\usepackage{subfloat}
\pdfoutput = 1
\usepackage{graphics}
\usepackage{graphicx} 
\usepackage{graphicx} 
\usepackage{verbatim} 
\usepackage{amsmath} 
\usepackage{multirow}
\usepackage{hhline}
\begin{document}
\date{Today}
\title{{\bf{\Large Noncommutative effects on holographic superconductors with power Maxwell electrodynamics}}}

\author{
{\bf {\normalsize Suchetana Pal}$^{a}$
\thanks{suchetanapal92@gmail.com, sp15rs004@iiserkol.ac.in}},\,
{\bf {\normalsize Sunandan Gangopadhyay}$^{a,b}
$\thanks{sunandan.gangopadhyay@gmail.com, sunandan@iiserkol.ac.in,  sunandan@associates.iucaa.in}}\\
$^{a}$ {\normalsize Indian Institute of Science Education and Research, Kolkata}\\{\normalsize Mohanpur, Nadia 741246, India}\\[0.2cm]
$^{b}${\normalsize Visiting Associate in Inter University Centre for Astronomy \& Astrophysics,}\\
{\normalsize Pune 411007, India}\\[0.1cm]
}
\date{}

\maketitle

\begin{abstract}
{\noindent The matching method is employed to analytically investigate the properties of holographic superconductors in higher dimensions in the framework of power Maxwell electrodynamics taking into account the effects of spacetime noncommutativity. The relationship between the critical temperature and the charge density and the value of the condensation operator is obtained first. The Meissner like effect is then studied. The analysis indicate that larger values of the noncommutative parameter and the parameter $q$ appearing in the power Maxwell theory makes the condensate difficult to form. The critical magnetic field however increases with increase in the noncommutative parameter $\theta$.  }
\end{abstract}
\vskip 1cm

\section{Introduction}

It is well known that the BCS theory of superconductivity \cite{bcs} has proved to be an important theoretical discovery which describes the properties of low temperature superconductors. However, there has been no successful theory yet to understand the superconductivity in a class of materials that exhibit the phenomenon at high temperatures. Examples of such materials are the high $T_c$ cuprates. It is reasonably evident that these are strongly coupled materials and hence lies the difficulty in developing a theory for them.

Recently, there has been an upsurge in the study of holographic superconductors. The importance of these were realized as they reproduced some properties of high $T_c$ superconductors. The theoretical input that goes in the construction of these holographic superconductor models is the gauge/gravity correspondence  first discovered in the context of string theory \cite{adscft1}-\cite{polyakov}. The physical mechanism behind this understanding involves the demonstration of a spontaneous symmetry breaking in an Abelian Higgs model in a black hole background that is asymptotically AdS \cite{gub}-\cite{prl}. Thereafter, several important properties of these gravity models have been studied analytically
\cite{rob}-\cite{dghorai}.

An important aspect of superconductors is their response to an external magnetic field known as the Meissner effect \cite{tink}. The response shows perfect diamagnetism as the temperature is lowered below a critical temperature $T_c$.
Quite a few number of investigations have been carried out both numerically \cite{maeda}-\cite{john} as well as analytically \cite{royc}-\cite{sg}. However, these studies were restricted mainly to the framework of Maxwell electrodynamics. In \cite{dib}, the conventional action for Maxwell electrodynamics was replaced by the power Maxwell action \cite{her}. The motivation for such a study comes from the question of investigating the behaviour of the condensate in the presence of higher curvature corrections coming from the power Maxwell theory.

Noncommutativity of spacetime is another important area of theoretical physics where considerable research has been carried out. The idea which first came in 1947 \cite{snyder} was brought into forefront from studies in string theory \cite{seiberg}. Very recently, black hole backgrounds have been provided incorporating the ideas of noncommutativity \cite{nic1}-\cite{rbsg}. The noncommutative effect gets introduced here by a smeared source of matter. This is then used to solve Einstein's equation of general relativity.

In this paper, we want to investigate the role of noncommutative spacetime on the properties of holographic superconductors. Such an investigation has been carried out earlier in \cite{ghorai1} in the framework of Born-Infeld electrodynamics. Here, we shall carry out such a study in power Maxwell electrodynamics using the matching method approach \cite{soda} in the probe limit approximation which essentially means that the backreaction of the spacetime has been neglected. Moreover, we shall also investigate the Meissner effect in this framework and study the role played by spacetime noncommutativity.

The paper is organized as follows. In section 2, the basic formalism for the $d$-dimensional holographic superconductor in noncommutative spacetime coupled to power Maxwell electrodynamics is presented. In section 3, we obtain the relationship between the critical temperature and the charge density and the value of the condensation operator using the matching method approach. In section 4, we investigate the Meissner like effect using the same approach. 
We finally conclude in section 5.

\section{Basic analytical set up}
The analysis proceeds by setting up a gravitational dual for higher dimensional superconductors. 
We consider the metric of a $d$-dimensional planar noncommutative Schwarzschild-AdS spacetime as the gravity dual of the holographic superconductor. The metric of such a black hole spacetime reads
\begin{equation} 
ds^{2}= -f\left ( r \right )dt^{2}+\frac{1}{f\left ( r \right )}dr^{2}+r^{2}dx_{i}dx^{i}
\label{s1}
\end{equation}
where $f(r)$ is given by \cite{nic1}
\begin{equation}
f(r)=\frac{r^{2}}{L^{2}}-\frac{2MG_{d}}{r^{d-3}\Gamma(\frac{d-1}{2})}\gamma\left(\frac{d-1}{2},\frac{r^{2}}{4\theta}\right)+ k
\label{eq02}
\end{equation}

\noindent and $\gamma(s,x)$ is the lower incomplete gamma function given by
\begin{equation}
\gamma(s,x)=\int_{0}^{x}t^{s-1}e^{-t} dt
\label{eq03}
\end{equation}
and $k$ denotes the curvature. $dx^{i}dx_{i}$ represents the line element of a $(d-2)$-dimensional hypersurface with vanishing curvature. In this analysis, we shall set $k=0$ since we require a planar holographic superconductor. The metric therefore takes the form 
\begin{equation}
f(r)=\frac{r^{2}}{L^{2}}- \frac{2MG_{d}}{r^{d-3}\Gamma \left ( \frac{d-1}{2} \right )}\gamma\left(\frac{d-1}{2},\frac{r^{2}}{4\theta}\right)
\label{eq04}~.
\end{equation}
The horizon radius can be obtained by setting $[f(r)]_{r=r_{+}}=0$ :
\begin{equation}
r_{+}^{d-1}=
\frac{2MG_{d}}{\Gamma(\frac{d-1}{2})}\gamma\left(\frac{d-1}{2},\frac{r_{+}^{2}}{4\theta}\right)
\label{eq05}
\end{equation}

\noindent where we have set $L=1$ for convenience. Using eq.(\ref{eq05}),
eq.(\ref{eq04}) can be recast as

\begin{equation}
f(r) = r^{2}\left(1- \frac{r_{+}^{d-1}}{r^{d-1}}\frac{\gamma(\frac{d-1}{2},\frac{r^{2}}{4\theta})}{\gamma(\frac{d-1}{2},\frac{r_{+}^{2}}{4\theta})}\right).
\label{eq06}
\end{equation}

\noindent The Hawking temperature $T$ of the black hole is given by
\begin{equation}
T= \frac{f^{'}(r_{+})}{4\pi}~.
\label{eq07}
\end{equation}
By the gauge/gravity duality, this temperature is interpreted as the temperature of the boundary field theory. Using eq.(s) (\ref{eq04}) and (\ref{eq05}), we get 

\begin{eqnarray}
T & = & \frac{1}{4\pi}\left[(d-1)r_{+}-r_{+}^{2}\frac{\gamma^{'}(\frac{d-1}{2},\frac{r_{+}^{2}}{4\theta})}{\gamma(\frac{d-1}{2},\frac{r_{+}^{2}}{4\theta})}\right]\nonumber\\
&=&\frac{r_{+}}{4\pi}\left[d-1-\frac{4MG_{d}}{\Gamma(\frac{d-1}{2})}\frac{e^{-\frac{r_{+}^{2}}{4\theta}}}{(4\theta)^{\frac{d-1}{2}}}\right]~.
\label{eq08}
\end{eqnarray}

\noindent In the limit $\theta\rightarrow0$, the coefficient of the $t$-component of the noncommutative metric reduces to

\begin{equation}
f(r)= r^{2}\left(1-\frac{r_{+}^{d-1}}{r^{d-1}}\right)
\label{eq09}
\end{equation}
where $r_{+}= 2MG_{d}$. This metric is the planar Schwarzschild-AdS black hole in $d$-spacetime dimensions.

 
\noindent The Hawking temperature for this black hole reads 
\begin{equation}
T=\frac{(d-1)r_{+}}{4\pi}~.
\label{eq10}
\end{equation}

\noindent We now write down an appropriate action for the bulk which can explain the phase transition at the boundary. The action involves a gravity theory with a negative cosmological constant together with a complex scalar field $\psi$  minimally coupled to the power Maxwell field
\begin{eqnarray}
S=\int d^{d}x \sqrt{-g} \left[R-2\Lambda -\beta(F_{\mu\nu}F^{\mu\nu})^{q}-|\bigtriangledown_{\mu}\psi-iA_{\mu}\psi|^{2}-m^{2}|\psi|^{2}\right]
\label{eq11}
\end{eqnarray}
where $\beta$ is the coupling constant of power Maxwell electrodynamics, $q$ is the power parameter of the power Maxwell field, $\Lambda=-\frac{(d-1)(d-2)}{2}$ is the cosmological constant.

\noindent In the subsequent discussion, we shall work in the probe limit which means that the effect of back reaction of the matter fields on the metric is neglected.

\noindent The equations of motion for the Maxwell and the scalar field can be obtained by varying the action. These read 

\begin{eqnarray}
 \frac{4\beta q}{\sqrt{-g}}\partial_{\mu}(\sqrt{-g}(F_{\lambda\sigma}F^{\lambda\sigma})^{q-1} F^{\mu\nu})-i(\psi^{*}\partial^{\nu}\psi-\psi(\partial^{\nu}\psi)^{*})-2A^{\nu}|\psi|^{2}=0  
 \label{eq12}
\end{eqnarray}

\begin{eqnarray}
\partial_{\mu}(\sqrt{-g}\partial^{\mu}\psi)-i\sqrt{-g}A^{\mu}\partial_{\mu}\psi-i\partial_{\mu}(\sqrt{-g}A^{\mu}\psi)-\sqrt{-g}A_{\mu}A^{\mu}\psi-\sqrt{-g}m^{2}\psi=0~.
\label{eq13}
\end{eqnarray}
To make further progress, we make the following ansatz 

\begin{equation}
A=\phi(r)dt \quad ,\quad  \psi=\psi(r)~.
\label{eq14}
\end{equation}

\noindent With this ansatz, eq.(s)(\ref{eq12}) and (\ref{eq13}) now take the form
\begin{equation}
\partial_{r}^{2}\phi+\frac{1}{r}\left(\frac{d-2}{2q-1}\right)\partial_{r}\phi-\frac{2\phi \psi^{2}(\partial_{r}\phi)^{2(1-q)}}{(-1)^{q-1}2^{q+1}\beta q(2q-1)f(r)}=0
\label{eq15}
\end{equation}

\begin{equation}
\partial_{r}^{2}\psi +\left(\frac{f^{'}}{f}+ \frac{d-2}{r}\right)\partial_{r}\psi + \frac{\phi^{2}\psi}{f^{2}}-\frac{m^{2}\psi}{f}=0~.
\label{eq16}
\end{equation}

 \noindent Making a change of variable $z=\frac{r_{+}}{r}$, the above equations take the form 

\begin{eqnarray}
\partial_{z}^{2}\phi+\frac{1}{z}\left(2-\frac{d-2}{2q-1}\right)\partial_{z}\phi  + \frac{2\phi(z)\psi^{2}(z)r_{+}^{2q}(\partial_{z}\phi)^{2(1-q)}}{z^{4q}2^{q+1}(-1)^{3q}\beta q (2q-1)f(z)}=0
\label{eq17}
\end{eqnarray}
 
\begin{equation}
\partial_{z}^{2}\psi +\left(\frac{f^{'}(z)}{f(z)}- \frac{d-4}{z}\right)\partial_{z}\psi + \frac{\phi^{2}\psi r_{+}^{2}}{z^{4}f^{2}(z)}-\frac{m^{2}\psi r_{+}^{2}}{z^{4}f(z)}=0~.
\label{eq18}
\end{equation}

\noindent In the limit $ z\rightarrow0 $, the asymptotic behaviour of the fields $\phi$ and $\psi$ can be expressed as 

\begin{equation}
\phi(z)=\mu-\frac{\rho^{\frac{1}{2q-1}}}{r_{+}^{\frac{d-2}{2q-1}-1}}z^{\frac{d-2}{2q-1}-1}
\label{eq19}
\end{equation}

\begin{equation}
\psi(z)=\frac{\psi_{-}}{r_{+}^{\lambda_{-}}}z^{\lambda_{-}} +\frac{\psi_{+}}{r_{+}^{\lambda_{+}}}z^{\lambda_{+}} 
\label{eq20}
\end{equation}
where

\begin{equation}
\lambda_{\pm}=\frac{1}{2}\left[d-1\pm \sqrt{(d-1)^{2}+4m^{2}}\right]
\label{eq21}~.
\end{equation}

\noindent The gauge/gravity duality interprets $\mu$ and $\rho$ as the chemical potential and the charge density. $\psi_{+}$ and $\psi_{-}$ are the vacuum expectation values of the dual operator $O$. For the rest of our analysis, we shall choose $\psi_{+}=\left \langle O_{+} \right \rangle$ and $\psi_{-}=0$.


\section{Critical temperature, charge density relation and condensation operator}
To begin with, we start by substituting $z=\frac{r_{+}}{r}$ in eq.
(\ref{eq06}). This yields
 \begin{equation}
f(z)= \frac{r_{+}^{2}}{z^{2}}g_{0}(z)
\label{eq22}
\end{equation}

\noindent where
 \begin{equation}
g_{0}(z)=\left(1-z^{d-1}\frac{\gamma(\frac{d-1}{2},\frac{r_{+}^{2}}{4\theta z^{2}})}{\gamma(\frac{d-1}{2},\frac{r_{+}^{2}}{4\theta})}\right).
\label{eq23}
\end{equation}
We now apply the matching method to proceed further. Near the horizon, that is $z=1$, we make Taylor series expansions of the fields $\phi(z)$ and $\psi(z)$. These read

\begin{eqnarray}
\phi(z) & = & \phi(1)-\phi^{'}(1)(1-z)+\frac{1}{2}\phi^{''}(1)(1-z)^{2}+ O((1-z)^{3})\nonumber
\\& = & -\phi^{'}(1)(1-z)+\frac{1}{2}\phi^{''}(1)(1-z)^{2}+ O((1-z)^{3})   
\label{eq24}
\end{eqnarray}

\begin{eqnarray}
 \psi(z)=\psi(1)-\psi^{'}(1)(1-z)+\frac{1}{2}\psi^{''}(1)(1-z)^{2}+O((1-z)^{3})
\label{eq25}
\end{eqnarray}
since $\phi(1)=0$.

\noindent To evaluate the expressions for  $\phi^{''}(1)$ and
 $\psi^{'}(1)$, $\psi^{''}(1)$, we look at eq.(s)
 (\ref{eq17}) and (\ref{eq18}) for $z=1$. This yields
 
\begin{eqnarray}
\phi^{''}(1) & = & \left(\frac{d-2}{2q-1}-2\right)\phi^{'}(1)-\frac{2r_{+}^{2(q-1)}\psi^{2}(1)(\phi^{'}(1))^{3-2q}}{2^{q+1}(-1)^{3q}(\beta q)(2q-1)g_{0}^{'}(1)} \label{eq26}
\\\psi^{'}(1) & = &\frac{m^{2}}{g_{0}^{'}(1)}\psi(1)
 \label{eq27}
\\\psi^{''}(1) & = & \frac{1}{2}\left[d-4-\frac{g_{0}^{''}(1)}{g_{0}^{'}(1)}+\frac{m^{2}}{g_{0}^{'}(1)}\right]\frac{m^{2}}{g_{0}^{'}(1)}\psi(1)-\frac{\phi^{'2}(1)\psi(1)}{2r_{+}^{2}g_{0}^{'2}(1)}~.
 \label{eq28}
\end{eqnarray}
Substituting these expressions in eq.(s) (\ref{eq24}) and (\ref{eq25}), we get
 
 
\begin{multline}
\phi(z)
 =-\phi^{'}(1)(1-z)\\+\frac{1}{2}(1-z)^{2}\left[\left(\frac{d-2}{2q-1}-2\right)\phi^{'}(1)-\frac{2r_{+}^{2(q-1)}\psi^{2}(1)(\phi^{'}(1))^{3-2q}}{(-1)^{3q}2^{q+1}(\beta q)(2q-1)g_{0}^{'}(1)}\right]
\label{eq29}
\end{multline}

 
\begin{multline}
  \psi(z)
 =\psi(1)-\frac{m^{2}}{g_{0}^{'}(1)}\psi(1)(1-z)\\
 +\frac{1}{2}(1-z)^{2}\left[\frac{1}{2}\left(d-4-\frac{g_{0}^{''}(1)}{g_{0}^{'}(1)}+\frac{m^{2}}{g_{0}^{'}(1)}\right)\frac{m^{2}}{g_{0}^{'}(1)}\psi(1)-\frac{\phi^{'2}(1)\psi(1)}{2r_{+}^{2}g_{0}^{'2}(1)} \right]~.
\label{eq30}
\end{multline}

\noindent We now proceed to implement the matching method. We match the above solutions with the asymptotic solutions (\ref{eq19}) and (\ref{eq20}) at $z=z_{m}$. This gives the following relations

  
\begin{multline}
 \mu -\frac{\rho^{\frac{1}{2q-1}}z_{m}^{\frac{d-2}{2q-1}-1}}{(r_{+})^{\frac{d-2}{2q-1}-1}}
 =\\v(1-z_{m})+\frac{1}{2}(1-z_{m})^{2}\left[\left(2-\frac{d-2}{2q-1}\right)v-\frac{2r_{+}^{2(q-1)}\alpha^{2}(-v)^{3-2q}}{(-1)^{3q}2^{q+1}(\beta q)(2q-1)(g_{0}^{'})}\right]
\label{eq31}
\end{multline}


\begin{multline}
 \frac{\left \langle O_{+} \right \rangle z_{m}^{\lambda_{+}}}{r_{+}^{\lambda_{+}}}=\\\alpha - (1-z_{m})\alpha\left(\frac{m^{2}}{g_{0}^{'}(1)} \right )+ \frac{1}{2}\alpha(1-z_{m})^{2}
 \left[\frac{1}{2}\left(\frac{m^{2}}{g_{0}^{'}(1)}\right)
 \left(d-4-\frac{g_{0}^{''}(1)}{g_{0}^{'}(1)}+\frac{m^{2}}{g_{0}^{'}(1)}\right)-\frac{\tilde{v}^{2}}{2g_{0}^{'}(1)^{2}}\right]~.
\label{eq32}
\end{multline}
where $v=-\phi^{'}(1), \alpha=\psi(1)$ and $ \tilde{v}=\frac{v}{r} $.

\noindent Taking derivative on both sides of eq.(\ref{eq31}) and (\ref{eq32}) yields
\begin{multline}
  -\frac{\rho^{\frac{1}{2q-1}}z_{m}^{\frac{d-2}{2q-1}-2}}{(r_{+})^{\frac{d-2}{2q-1}-1}}\left(\frac{d-2}{2q-1}-1\right)
 =\\-v-(1-z_{m})\left[\left(2-\frac{d-2}{2q-1}\right)v-\frac{2r_{+}^{2(q-1)}\alpha^{2}(-v)^{3-2q}}{(-1)^{3q}2^{q+1}(\beta q)(2q-1)g_{0}^{'}(1)}\right]
\label{eq33}
\end{multline}
\begin{multline}
 \lambda_{+}\frac{\left \langle O_{+} \right \rangle z_{m}^{\lambda_{+}-1}}{r_{+}^{\lambda_{+}}}=\\\alpha\left(\frac{m^{2}}{g_{0}^{'}(1)}\right)
  -\alpha(1-z_{m})\left[\frac{1}{2}\left(\frac{m^{2}}{g_{0}^{'}(1)}\right)
 \left(d-4-\frac{g_{0}^{''}(1)}{g_{0}^{'}(1)}+\frac{m^{2}}{g_{0}^{'}(1)}\right)-\frac{\tilde{v}^{2}}{2g_{0}^{'}(1)^{2}}\right].
\label{eq34}
\end{multline}
From the above set of equations together with eq.(\ref{eq08}), it is simple to obtain
\begin{multline}
\alpha^{2}\equiv\alpha^{2}_{NC}=- \frac{(-1)^{5q-3}2^{q}(\beta q)(2q-1)g_{0}^{'}(1)}{\tilde{v}_{NC}^{2(1-q)}(1-z_{m})}\\
\times \left[1+\left(2-\frac{d-2}{2q-1}\right)(1-z_{m}) \right]\times\left(\frac{(T_{c})_{NC}}{T}\right)^{\frac{d-2}{2q-1}}\left[1-\left(\frac{T}{(T_{c})_{NC}}\right)^{\frac{d-2}{2q-1}}\right]
\label{eq35}
\end{multline}
where
\begin{eqnarray}
(T_{c})_{NC} &=&\xi_{NC}\rho^{\frac{1}{d-2}}\\
\xi_{NC} &=&-\frac{z_{m}^{(\frac{d-2}{2q-1}-2)(\frac{2q-1}{d-2})}}{\tilde{v}_{NC}^{\frac{2q-1}{d-2}}}\left(\frac{g_{0}^{'}(1)}{4\pi}\right)\frac{(\frac{d-2}{2q-1}-1)^{\frac{2q-1}{d-2}}}{[1+(2-\frac{d-2}{2q-1})(1-z_{m})]^{\frac{2q-1}{d-2}}}~.
\end{eqnarray}
Note that $NC$ in the above equations stand for the noncommutative case. The above results give the relation between the critical temperature and the charge density. It can be observed from the analytical results that the critical temperature decreases with increase in the noncommutative parameter $\theta$ which clearly indicate that the condensate gets harder to form as the spacetime noncommutativity increases. However, as the mass of the black hole increases, the critical temperature for a particular value of $\theta$ increases which tells that the effects of spacetime noncommutativity becomes prominent for lower mass black holes.
Further, we can infer from the Tables 3 and 4 (comparing the results with Table 2) that the onset of power Maxwell electrodynamics (for a value of $q\neq 1$) makes the condensate difficult to form. However, in this case also the effect of the power Maxwell theory on the formation of the condensate decreases with increase in the mass of the black holes.\\

\noindent In the Tables 1, 2, 3 and 4, we present the analytical results for $\xi_{NC}$ for different values of $M$ and $\theta$.
\begin{table}[ht]
\caption{Analytical values of $\xi_{NC}$ for different values of $M$ and $\theta$ [$q=1$, $m^{2}=0$, $z_{m}=0.5$ and $d=5$] }   
\centering                          
\begin{tabular}{|c| c| c| c| }            
\hline
 $\theta$ & \multicolumn{3}{c|}{$\xi_{NC}$}  \\
\hline
 & $MG_{d}=10$ & $MG_{d}=50$ & $MG_{d}=100$  \\
\hline
0.3 & 0.1507 & 0.16933 & 0.1702  \\ 
\hline
0.5 & 0.1384 & 0.16058 & 0.1678  \\ 
\hline
0.7 & 0.1395 & 0.1492 & 0.1608  \\ 
\hline 
0.9 & 0.1439 & 0.1418 & 0.1525  \\ 
\hline 
\end{tabular}
\label{t3}  
\end{table}


\begin{table}[ht]
\caption{Analytical values of $\xi_{NC}$ for different values of $M$ and $\theta$ [$q=1$, $m^{2}=-3$, $z_{m}=0.5$ and $d=5$] }   
\centering                          
\begin{tabular}{|c| c| c| c| }            
\hline
 $\theta$ & \multicolumn{3}{c|}{$\xi_{NC}$}  \\
\hline
 & $MG_{d}=10$ & $MG_{d}=50$ & $MG_{d}=100$  \\
\hline
0.3 & 0.1761 & 0.2003 & 0.2015  \\ 
\hline
0.5 & 0.1649 & 0.18798 & 0.1977  \\ 
\hline
0.7 & 0.1701 & 0.1744 & 0.1883  \\ 
\hline 
0.9 & 0.1767 & 0.1669 & 0.1782  \\ 
\hline 
\end{tabular}
\label{t3a}  
\end{table}



\begin{table}[ht]
\caption{Analytical values of $\xi_{NC}$ for different values of $M$ and $\theta$ [ $q=5/4$, $m^{2}=-3$, $z_{m}=0.5$ and $d=5$] }   
\centering                          
\begin{tabular}{|c| c| c| c| }            
\hline
 $\theta$ & \multicolumn{3}{c|}{$\xi_{NC}$}  \\
\hline
 & $MG_{d}=10$ & $MG_{d}=50$ & $MG_{d}=100$  \\
\hline
0.3 & 0.1015 & 0.1126 & 0.1134  \\ 
\hline
0.5 & 0.0981 & 0.1067 & 0.1114  \\ 
\hline
0.7 & 0.1020 & 0.1008 & 0.1069  \\ 
\hline 
0.9 & 0.1056 & 0.0980 & 0.1024  \\ 
\hline 
\end{tabular}
\label{t3b}  
\end{table}


\begin{table}[ht]
\caption{Analytical values of $\xi_{NC}$ for different values of $M$ and $\theta$ [ $q=7/4$, $m^{2}=-3$, $z_{m}=0.5$ and $d=5$] }   
\centering                          
\begin{tabular}{|c| c| c| c| }            
\hline
 $\theta$ & \multicolumn{3}{c|}{$\xi_{NC}$}  \\
\hline
 & $MG_{d}=10$ & $MG_{d}=50$ & $MG_{d}=100$  \\
\hline
0.3 & 0.0167 & 0.0177 & 0.0178  \\ 
\hline
0.5 & 0.0172 & 0.0171 & 0.0176  \\ 
\hline
0.7 & 0.0183 & 0.0167 & 0.0171  \\ 
\hline 
0.9 & 0.0187 & 0.0168 & 0.0168  \\ 
\hline 
\end{tabular}
\label{t3c}  
\end{table}

\noindent From eq.(s).(\ref{eq32}) and (\ref{eq34}), we obtain the expression for $\tilde{v}_{NC}$


\begin{multline}
\tilde{v}\equiv\tilde{v}_{NC}^{2}=
m^{4}+m^{2}g_{0}^{'}(1)
\left(d-4-\frac{g_{0}^{''}(1)}{g_{0}^{'}(1)}\right)-
\left(\frac{4m^{2}g_{0}^{'}(1)(z_{m}+\lambda_{+}(1-z_{m}))}{(1-z_{m})(2z_{m}+\lambda_{+}(1-z_{m}))}\right)\\ + \left(\frac{4g_{0}^{'2}(1)\lambda_{+}}{(1-z_{m})(2z_{m}+\lambda_{+}(1-z_{m}))}\right)~.
\label{eq38}
\end{multline}

\noindent Now we have all the expressions in hand required to compute the condensation operator for this problem. The expression for $\left \langle O_{+} \right \rangle$ can be obtained by substituting $\tilde{v}_{NC}$ from eq. (\ref{eq38}) in eq.(\ref{eq33}). This gives


\begin{equation}
\left \langle O_{+} \right \rangle_{NC}=\frac{r_{+}^{\lambda_{+}}(1-\frac{m^{2}(1-z_{m})}{2g_{0}^{'}(1)})}{z_{m}^{\lambda_{+}}(1+\frac{\lambda_{+}(1-z_{m})}{2z_{m}})}\sqrt{\mathit{A}_{NC}}\times
\left(\frac{(T_{c})_{NC}}{T}\right)^{\frac{d-2}{2(2q-1)}}
\sqrt{1-\left(\frac{T}{(T_{c})_{NC}}\right)^{\frac{d-2}{2q-1}}}
\label{eq39}
\end{equation}

\noindent where

\begin{equation}
\mathit{A}_{NC}=\frac{(-1)^{5q-3}2^{q}(\beta q)(2q-1)(-g_{0}^{'}(1))}{\tilde{v}_{NC}^{2(1-q)}(1-z_{m})}\times\left[1+\left(2-\frac{d-2}{2q-1}\right)(1-z_{m})\right]~.
\label{eq40}
\end{equation}
Now we proceed to take the $\theta \rightarrow 0$ limit of the above findings.
The expression for $\alpha\equiv\alpha_{C}=\psi(1)$ now reduces to\footnote{It is to be noted that our expression differs from that in \cite{dib} since an algebraic error was made in that paper.}

\begin{multline}
\alpha^{2}_{C}= \frac{(-1)^{5q-1}2^{q}(\beta q)(2q-1)(d-1)}{\tilde{v}^{2(1-q)}(1-z_{m})}\\
\times \left[1+\left(2-\frac{d-2}{2q-1}\right)(1-z_{m}) \right]\times\left(\frac{(T_{c})_{C}}{T}\right)^{\frac{d-2}{2q-1}}\left[1-\left(\frac{T}{(T_{c})_{C}}\right)^{\frac{d-2}{2q-1}}\right]
\label{eq41}
\end{multline}
where
\begin{equation}
(T_{c})_{C}=\xi_{C}\rho^{\frac{1}{d-2}}
\label{eq42}
\end{equation}
with $\xi_{C}$ given by
\begin{equation}
\xi_{C}=\frac{z_{m}^{(\frac{d-2}{2q-1}-2)(\frac{2q-1}{d-2})}}{\tilde{v}_{C}^{\frac{2q-1}{d-2}}}\left(\frac{d-1}{4\pi}\right)\frac{(\frac{d-2}{2q-1}-1)^{\frac{2q-1}{d-2}}}{[1+(2-\frac{d-2}{2q-1})(1-z_{m})]^{\frac{2q-1}{d-2}}}~.
\label{eq43}
\end{equation}
The expression for $\tilde{v}\equiv\tilde{v}_{C}$ and the condensation operator in the commutative case take the form
\begin{multline}
\tilde{v}_{C}=
\left
(m^{4}+2m^{2}(d-1)\left[\frac{2(z_{m}+\lambda_{+}(1-z_{m}))}{(1-z_{m})(2z_{m}+\lambda_{+}(1-z_{m}))}+1\right]+\frac{4\lambda_{+}(d-1)^{2}}{(1-z_{m})(2z_{m}+\lambda_{+}(1-z_{m}))}
\right)^{\frac{1}{2}}
\label{eq44}
\end{multline}

\begin{equation}
\left \langle O_{+} \right \rangle_{C}=\frac{r_{+}^{\lambda_{+}}(1+\frac{m^{2}(1-z_{m})}{2(d-1)})}{z_{m}^{\lambda_{+}}(1+\frac{\lambda_{+}(1-z_{m})}{2z_{m}})}\sqrt{\mathit{A}_{C}}\times
\left(\frac{(T_{c})_{C}}{T}\right)^{\frac{d-2}{2(2q-1)}}
\sqrt{1-\left(\frac{T}{(T_{c})_{C}}\right)^{\frac{d-2}{2q-1}}}
\label{eq45}
\end{equation}

where
\begin{equation}
\mathit{A}_{C}=\frac{(-1)^{5q-3}2^{q}(\beta q)(2q-1)(d-1)}{\tilde{v}_{C}^{2(1-q)}(1-z_{m})}\times\left[1+\left(2-\frac{d-2}{2q-1}\right)(1-z_{m})\right]~.
\label{eq46}
\end{equation}


\noindent In Table 5, we display the analytical results for $\xi_{C}$
for the commutative case.
\begin{table}[ht]
\caption{Analytical values of $\xi_{C}$ for $z_{m}=0.5$ and $d=5$}   
\centering                          
\begin{tabular}{|c| c| c| c|  }            
\hline
$m^2$ & \multicolumn{3}{c|}{$\xi_{C}$}  \\
\hline
 &$q=1$ & $q=5/4$ & $q=7/4$   \\
\hline
 0 & 0.17028 &0.0880 &0.0117 \\ 
\hline
 -3 & 0.2017 & 0.1135 &0.0179 \\ 
\hline
\end{tabular}
\label{t5}  
\end{table}
In Figure 5, we use the analytical results in Tables 2, 3 and 4 to plot $\xi$ vs $\theta$ for different values of the power Maxwell parameter. 

\begin{figure}[h!]
\centering
\includegraphics[width=8cm]{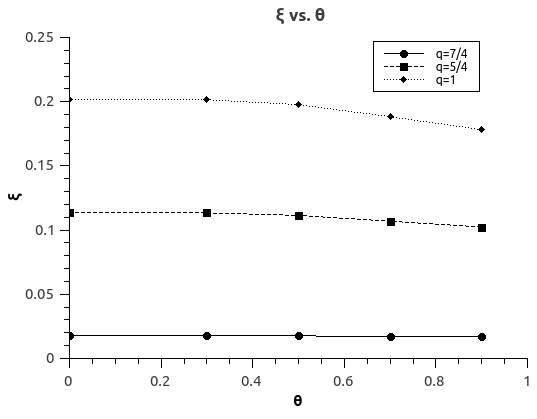}

\caption{$\xi$ vs $\theta$ plot : $z_{m}=0.5, MG_{d}=100, m^2=-3, d=5$}
\label{fig1}
\end{figure}


\section{Meissner like effect}
In this section we introduce an external magnetic field $B$ in the bulk theory and observe how the condensation behaves at low temperature for noncommutative black hole background in the bulk. The intention is to find a critical magnetic filed $B_{c}$ above which the condensation vanishes. We therefore make the following ansatz
\begin{equation}
A_{t}=\phi(z) \quad, \quad  A_{y}=Bx \quad, \quad \psi=\psi(x,z)~.
\label{eq47}
\end{equation} 
The equation of motion for the complex scalar field  $\psi$ that follows from the above ansatz reads
\begin{multline}
\partial_{z}^{2}\psi(x,z)+\left(\frac{f^{'}(z)}{f(z)}-\frac{d-4}{z}\right)\partial_{z}\psi(x,z)\\+\frac{\phi^{2}(z)\psi(x,z) r_{+}^{2}}{z^{4}f^{2}(z)}-\frac{m^{2}r_{+}^{2}\psi(x,z)}{z^{4}f(z)}+\frac{1}{z^{2}f(z)}(\partial_{x}^{2}\psi-B^{2}x^{2}\psi)=0~.
\label{eq00}
\end{multline}
For solving the above equation, we write $\psi(x,z)$ as
\begin{equation}
\psi(x,z)=X(x)R(z)~.
\label{eq49}
\end{equation}
Substituting eq.(\ref{eq49}) in eq.(\ref{eq00}), we arrive at the following expression

\begin{multline}
\frac{z^{2}f(z)}{R(z)}\left[\partial_{z}^{2}R(z)+\left(\frac{f^{'}(z)}{f(z)}-\frac{d-4}{z}\right)\partial_{z}R(z)\right]
+\frac{\phi^{2}(z)r_{+}^{2}}{z^{2}f(z)}-\frac{m^{2}r_{+}^{2}}{z^{2}}
\\-\frac{1}{X(x)}\left[-\partial_{x}^{2}X(x)+B^{2}x^{2}X(x)\right]=0~.
\label{eq50}
\end{multline}
This equation implies that $X(x)$ satisfies a $1$-dimensional simple harmonic oscillator equation with frequency $B$

\begin{equation}
-X^{''}(x)+B^{2}x^{2}X(x)=\lambda_{n}BX(x)
\label{eq51}
\end{equation}
where the separation constant is given by $\lambda_{n}=2n+1$. For the rest of our analysis we shall set $n=0$, since this corresponds to the most stable mode.
\noindent The equation for $R(z)$ has the following form

\begin{equation}
R^{''}(z)+\left(\frac{f^{'}(z)}{f(z)}-\frac{d-4}{z}\right)R^{'}(z)+
\frac{\phi^{2}(z)r_{+}^{2}R(z)}{z^{4}f^{2}(z)}-\frac{m^{2}r_{+}^{2}R(z)}{z^{4}f(z)}\\=\frac{BR(z)}{z^{2}f(z)}~.
\label{eq52}
\end{equation}
Now we shall expand $R(z)$ in a Talylor series around $z=1$ and equate it with the asymptotic solution of $R(z)$ at some point $z=z_{m}$. 

\noindent The Taylor series expansion of $R(z)$ around $z=1$ reads

\begin{equation}
R(z)=R(1)-R^{'}(1)(1-z)+\frac{1}{2}R^{''}(1)(1-z)^{2}+O\left((1-z)^{3}\right)~.
\label{eq53}
\end{equation}

\noindent Further the asymptotic form for $R(z)$ reads

\begin{equation}
R(z)=\frac{\left \langle O \right \rangle_{+}}{r_{+}^{\lambda_{+}}}z^{\lambda_{+}}~.
\label{eq54}
\end{equation}

\noindent Equating these at $z=z_{m}$ yields
\begin{equation}
\left[\frac{\left \langle O \right \rangle_{+}}{r_{+}^{\lambda_{+}}}z^{\lambda_{+}}\right]_{z=z_{m}}=\left[R(1)-R^{'}(1)(1-z)+\frac{1}{2}R^{''}(1)(1-z)^{2}+O\left((1-z)^{3}\right)\right]_{z=z_{m}}
\label{eq55}~.
\end{equation}

\noindent Differentiating eq.(s)(\ref{eq53}) and (\ref{eq54}) with respect to $z$ and evaluating at $z=z_{m}$ yields 
\begin{equation}
\left[\lambda_{+}\frac{\left \langle O \right \rangle_{+}}{r_{+}^{\lambda_{+}}}z^{\lambda_{+}-1}\right]_{z=z_{m}}
=\left[R^{'}(1)-R{''}(1)(1-z)+O\left((1-z)^{3}\right)\right]_{z=z_{m}}~.
\label{eq56}
\end{equation}

\noindent Now for the noncommutative black hole spacetime (\ref{eq06}), we have from eq.(\ref{eq52})
\begin{equation}
R^{'}(1)=\left(\frac{m^{2}}{g_{0}^{'}(1)}+\frac{B}{r_{+}^{2}g_{0}^{'}(1)}\right)R(1)
\label{eq57}
\end{equation}

\begin{multline}
R^{''}(1)=
\frac{1}{2}\left[d-4+\frac{m^{2}}{g_{0}^{'}(1)}+\frac{B}{r_{+}^{2}g_{0}^{'}(1)}-\frac{g_{0}^{''}(1)}{g_{0}^{'}(1)}\right]\left[\frac{m^{2}}{g_{0}^{'}(1)}+\frac{B}{r_{+}^{2}g_{0}^{'}(1)}\right]R(1)
\\+\frac{BR(1)}{r_{+}^{2}g_{0}^{'}(1)}
-\frac{\phi^{'2}(1)R(1)}{2r_{+}^{2}g_{0}^{'2}(1)}~.
\label{eq58}
\end{multline}

\noindent Substituting $R^{'}(1)$ and $R^{''}(1)$ in eq.(s)(\ref{eq55}) and (\ref{eq56}), we have 

\begin{multline}
\left[\frac{\left \langle O \right \rangle_{+}}{r_{+}^{\lambda_{+}}}z_{m}^{\lambda_{+}}\right]=
R(1)
-\left(\frac{m^{2}}{g_{0}^{'}(1)}+\frac{B}{r_{+}^{2}g_{0}^{'}(1)}\right)(1-z_{m})R(1)\\
+\frac{1}{2}(1-z_{m})^{2}
[\frac{1}{2}\left[d-4+\frac{m^{2}}{g_{0}^{'}(1)}+\frac{B}{r_{+}^{2}g_{0}^{'}(1)}-\frac{g_{0}^{''}(1)}{g_{0}^{'}(1)}\right]\left[\frac{m^{2}}{g_{0}^{'}(1)}+\frac{B}{r_{+}^{2}g_{0}^{'}(1)}\right]R(1)
\\+\frac{BR(1)}{r_{+}^{2}g_{0}^{'}(1)}
-\frac{\phi^{'2}(1)R(1)}{2r_{+}^{2}g_{0}^{'2}(1)}]
\label{eq59}
\end{multline}

\begin{multline}
\left[\lambda_{+}\frac{\left \langle O \right \rangle_{+}}{r_{+}^{\lambda_{+}}}z_{m}^{\lambda_{+}-1}\right]=
\left(\frac{m^{2}}{g_{0}^{'}(1)}+\frac{B}{r_{+}^{2}g_{0}^{'}(1)}\right)R(1)\\
-(1-z_{m})
[\frac{1}{2}\left[d-4+\frac{m^{2}}{g_{0}^{'}(1)}+\frac{B}{r_{+}^{2}g_{0}^{'}(1)}-\frac{g_{0}^{''}(1)}{g_{0}^{'}(1)}\right]\left[\frac{m^{2}}{g_{0}^{'}(1)}+\frac{B}{r_{+}^{2}g_{0}^{'}(1)}\right]R(1)
\\+\frac{BR(1)}{r_{+}^{2}g_{0}^{'}(1)}
-\frac{\phi^{'2}(1)R(1)}{2r_{+}^{2}g_{0}^{'2}(1)}]~.
\label{eq60}
\end{multline}

\noindent Eq.(s)(\ref{eq59}) and (\ref{eq60}) yields a quadratic equation for $B$. This reads 
\begin{equation}
B^{2}+pr_{+}^{2}B+nr_{+}^{4}-\phi^{'2}(1)r_{+}^{2}=0
\label{eq61}
\end{equation}
where
\begin{equation}
p=2m^{2}+\left(d-4-\frac{g^{''}_{0}(1)}{g^{'}_{0}(1)}\right)g_{0}^{'}(1)+2g_{0}^{'}(1)-\frac{4g_{0}^{'}(1)(\lambda_{+}(1-z_{m})+z_{m})}{(1-z_{m})(\lambda_{+}(1-z_{m})+2z_{m})}
\label{eq62}
\end{equation}
and
 

\begin{multline}
n=m^{4}
+m^{2}g_{0}^{'}(1)
\left[\left(d-4-\frac{g^{''}_{0}(1)}{g^{'}_{0}(1)}\right)
-\frac{4(z_{m}+\lambda_{+}(1-z_{m}))}{(1-z_{m})(2z_{m}+\lambda_{+}(1-z_{m}))}\right]\\
+\frac{4\lambda{+}g_{0}^{'2}(1)}{(1-z_{m})(2z_{m}+\lambda_{+}(1-z_{m}))}~.
\label{eq63}
\end{multline}


\noindent Now when $B=B_{c}$, the condensate vanishes and hence we can take $\psi=0$ and eq.(\ref{eq17}) now takes the form 

\begin{equation}
\partial_{z}^{2}\phi+\frac{1}{z}\left(2-\frac{d-2}{2q-1}\right)\partial_{z}\phi= 0~.
\label{eq64}
\end{equation}
Solving this, we get 
\begin{equation}
\phi(z)=\left(\frac{\rho}{r_{+}^{d-2}}\right)^{\frac{1}{2q-1}}r_{+}(1-z^{\frac{d-2}{2q-1}-1})
\label{eq65}
\end{equation}
\begin{equation}
\Rightarrow \phi^{'2}(1)r_{+}^{2}=
\left(\frac{\rho}{r_{+}^{d-2}}\right)
^{\frac{2}{2q-1}}r_{+}^{4}
\left(\frac{d-2}{2q-1}-1\right)^{2}~.
\label{eq66}
\end{equation} 
Using eq.(\ref{eq66}) in eq.(\ref{eq61}) we get the expression for the critical magnetic field $B_{c}$ :
\begin{equation}
(B_{c})_{NC}=\frac{(-g_{0}^{'}(1))^{\frac{d-2}{2q-1}-2}}{2(4\pi)^{\frac{d-2}{2q-1}-2}\xi_{NC}^{\frac{d-2}{2q-1}}}(T_{c})_{NC}^{2}\times
\left[
\Omega_{NC}(d,q,m)-p\left(-\frac{4\pi\xi_{NC}}{g_{0}^{'}(1)}\right)^{\frac{d-2}{2q-1}}
\left(\frac{T}{(T_{c})_{NC}}\right)^{\frac{d-2}{2q-1}}\right]
\label{eq67}
\end{equation}
where

\begin{equation}
\Omega_{NC}(d,q,m)=
\left[4(\frac{d-2}{2q-1}-1)^{2}-(4n-p^{2})\left(-\frac{4\pi\xi_{NC}}{g_{0}^{'}(1)}\right)^{\frac{2(d-1)}{2q-1}}\left(\frac{T}{(T_{c})_{NC}}\right)^{\frac{2(d-1)}{2q-1}}\right]^{\frac{1}{2}}~.
\label{68}
\end{equation} 
Once again we take the $\theta\rightarrow 0$ limit of the above results. This gives the critical magnetic field in the commutative case :

\begin{equation}
(B_{c})_{C}=\frac{(d-1)^{\frac{d-2}{2q-1}-2}}{2(4\pi)^{\frac{d-2}{2q-1}-2}\xi_{c}^{\frac{d-2}{2q-1}}}(T_{c})_{C}^{2}\times
\left[
\Omega_{C}(d,q,m)-p\left(\frac{4\pi\xi_{C}}{d-1}\right)^{\frac{d-2}{2q-1}}
\left(\frac{T}{(T_{c})_{C}}\right)^{\frac{d-2}{2q-1}}\right]
\label{eq69}
\end{equation}
where
\begin{equation}
\Omega_{C}(d,q,m)=\left[4(\frac{d-2}{2q-1}-1)^{2}-(4n-p^{2})\left(\frac{4\pi\xi_{c}}{d-1}\right)^{\frac{2(d-1)}{2q-1}}\left(\frac{T}{(T_{c})_{C}}\right)^{\frac{2(d-1)}{2q-1}}\right]^{\frac{1}{2}}~.
\label{eq70}
\end{equation}
The above findings are displayed in Figures 2 and 3. It is evident from these figures that there exists a critical magnetic field as well as a critical temperature above which the superconducting phase vanishes. In Fig.2, we present our results for $B_{c}-T$ for two sets of values, namely, $m^{2}=-3$, $q=1$ and $d=5$ and $m^{2}=0$, $q=1$ and $d=5$ for different values of the noncommutative parameter $\theta$. In Fig.3, the plots are made for $q=5/4$ with $m^{2}=-3$ and  $d=5$ for different values of the noncommutative parameter $\theta$.

\begin{figure}[h!]
\centering
\includegraphics[width=8cm]{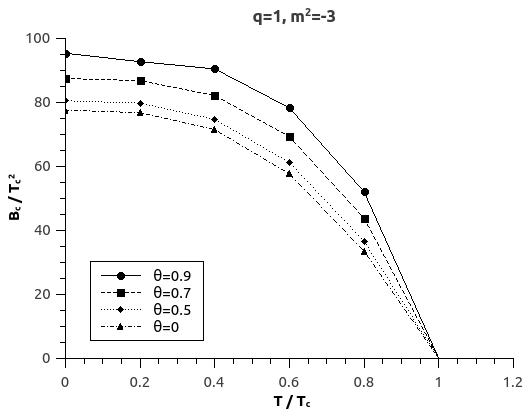}
\includegraphics[width=8cm]{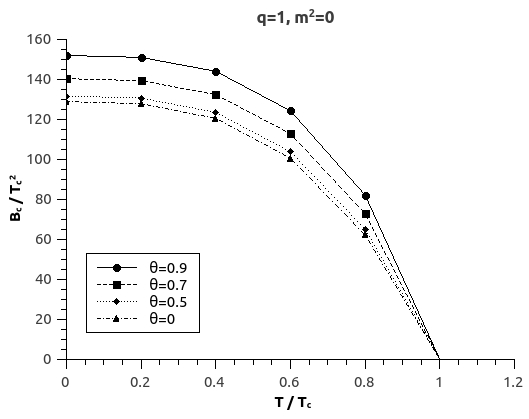}
\caption{$B_{c}/T_{c}^{2}$ vs $T/T_{c}$ plot : $z_{m}=0.5, MG_{d}=100, d=5$}
\label{fig1}
\end{figure}

\begin{figure}[h!]
\centering
\includegraphics[width=8cm]{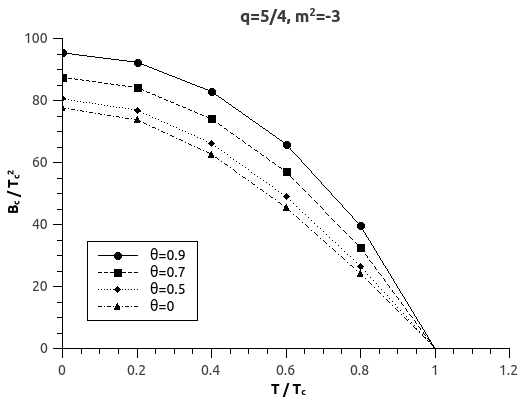}

\caption{$B_{c}/T_{c}^{2}$ vs $T/T_{c}$ plot : $z_{m}=0.5, MG_{d}=100, d=5$}
\label{fig1}
\end{figure}

\noindent It is interesting to note that the critical magnetic field above which the condensate vanishes increases with increase in the noncommutative parameter $\theta$.

\section{Conclusions}
In this paper, we have explored the role of noncommutativity of spacetime in holographic superconductors in the framework of power Maxwell electrodynamics
using the matching method technique. In our study, the relation between the critical temperature and the charge density has been obtained in $d$-dimensions. It is observed that the critical temperature not only depends on the charge density but also on the noncommutative parameter $\theta$, mass of the black hole and the parameter $q$ appearing in the power Maxwell theory. We have presented the analytical results for the ratio of the critical temperature and charge density for $d=5$. We have also analytically obtained the expression for the condensation operator in $d$-dimensions. Our analytical results indicate that the condensation gets harder to form in the presence of the power Maxwell parameter $q$ and the noncommutative parameter $\theta$. However, with increase in the mass of the black hole,
the critical temperature for a particular value of $\theta$ increases which reveals that the effects of spacetime noncommutativity becomes weaker for black holes with mass much larger in comparison to the noncommutative parameter $\theta$.
We also conclude from our results that the onset of power Maxwell electrodynamics (for a value of $q\neq 1$) makes the condensate harder to form. However, once again the effect of the power Maxwell parameter on the formation of the condensate decreases with increase in the mass of the black holes.
We then study the Meissner like effect by introducing an external magnetic field in our model. The critical magnetic field above which the condensate vanishes is obtained by the matching method and is observed to increase with increase in the noncommutative parameter $\theta$.

\section*{Acknowledgments} 
SP wants to thank the Council of Scientific and Industrial Research
(CSIR), Govt. of India for financial support.
SG acknowledges the support by DST SERB under Start Up Research Grant (Young Scientist), File No.YSS/2014/000180.



\end{document}